\documentclass[twocolumn,showpacs,prl,amsmath,amssymb]{revtex4}
\usepackage{graphicx}
\usepackage{dcolumn}
\usepackage{bm}

\begin{document}
\title{The Raman Fingerprint of Graphene}

\author{A. C. Ferrari$^1$}
\email{acf26@eng.cam.ac.uk}
\author{J. C. Meyer$^2$, V. Scardaci$^1$, C.
Casiraghi$^1$, M. Lazzeri$^2$, F. Mauri$^2$, S. Piscanec$^1$, Da
Jiang$^4$, K. S. Novoselov$^4$, S. Roth$^2$}
\author{A. K.
Geim$^4$}

\affiliation{$^1$Cambridge University, Engineering Department,
Trumpington Street, Cambridge CB3 0FA, UK\\
$^2$ Max Planck Institute for Solid State Research, Stuttgart 70569, Germany\\
$^3$Institut de Mineralogie et de Physique des Milieux Condenses,
Paris cedex 05, France\\ $^4$Department of Physics and Astronomy,
University of Manchester, Manchester, M13 9PL, UK}
\date{\today}

\begin{abstract}
Graphene is the two-dimensional (2d) building block for carbon
allotropes of every other dimensionality. It can be stacked into
3d graphite, rolled into 1d nanotubes, or wrapped into 0d
fullerenes. Its recent discovery in free state has finally
provided the possibility to study experimentally its electronic
and phonon properties. Here we show that graphene's electronic
structure is uniquely captured in its Raman spectrum that clearly
evolves with increasing number of layers. Raman fingerprints for
single-, bi- and few-layer graphene reflect changes in the
electronic structure and electron-phonon interactions and allow
unambiguous, high-throughput, non-destructive identification of
graphene layers, which is critically lacking in this emerging
research area.

\end{abstract}

\maketitle

The current interest in graphene can be attributed to three main
reasons. First, its electron transport is described by the Dirac
equation and this allows access to the rich and subtle physics of
quantum electrodynamics in a relatively simple condensed matter
experiment~\cite{ Novoselov2005proc, Novoselov2005nature,
Zhang2005nature, Novoselov2004, Novoselov2006}. Second, the
scalability of graphene devices to nano-dimensions~\cite{
Zhang2005apl, Peres2005, Berger2004, Wakabayashi2001, Nakada1996}
makes it a promising candidate for electronic applications,
because of its ballistic transport at room temperature combined
with chemical and mechanical stability. Remarkable properties
extend to bi-layer and few-layers graphene~\cite{ Novoselov2004,
Novoselov2006, Zhang2005apl, Berger2004, Scott2005}. Third,
various forms of graphite, nanotubes, buckyballs and others can
all be viewed as derivatives of graphene and, not surprisingly,
this basic material has been intensively investigated
theoretically for the past fifty years~\cite{ Wallace1947}. The
recent availability of graphene~\cite{ Novoselov2005proc} at last
allows to probe it experimentally, which paves the way to better
understanding the other allotropes and to resolve controversies.

Graphene samples can be obtained using the procedure of
Ref.~\cite{ Novoselov2005proc}, i.e. micro-mechanical cleavage of
graphite. Alternative procedures, such as exfoliation and growth,
so far only produced multi-layer samples~\cite{ Zhang2005apl,
Berger2004, Viculis2003}, but it is hoped that in the near future
efficient growth methods will be developed, as happened for
nanotubes. Despite the wide use of the micro-mechanical cleavage,
the identification and counting of graphene layers is a major
hurdle. Monolayers are a great minority amongst accompanying
thicker flakes. They cannot be seen in an optical microscope on
most substrates. Graphene layers only become visible when
deposited on the top of oxidized Si substrates with a finely tuned
thickness of the oxide layer (typically, 300 nm of SiO$_2$)
because, in this case, even a monolayer adds to the optical path
of reflected light to change the interference color with respect
to the empty substrate~\cite{ Novoselov2005proc, Novoselov2004}.
Atomic Force Microscopy (AFM) has been so far the only method to
identify single and few layers, but it is low throughput.
Moreover, due to the chemical contrast between graphene and the
substrate (which results in an apparent chemical thickness of
0.5-1nm, much bigger of what expected from the interlayer graphite
spacing~\cite{ Novoselov2005proc, Novoselov2004}), in practice, it
is only possible to distinguish between one and two layers by AFM
if films contain folds or wrinkles~\cite{ Novoselov2005proc,
Novoselov2004}. This poses a major limitation to the range of
substrates and is a set-back for the widespread utilization of
this material. Here, we show that graphene's electronic structure
is uniquely captured in its Raman spectrum. Raman fingerprints for
single-, bi- and few-layers reflect changes in the electronic
structure and allow unambiguous, high-throughput, non-destructive
identification of graphene layers, which is critically lacking in
this emerging research area.

The samples studied in this work were prepared by micromechanical
cleavage~\cite{Novoselov2005proc}. To provide the most definitive
identification of single and bi-layer graphene (beyond the layer
counting procedures by AFM) we perform Transmission Electron
Microscopy (TEM) on some of the samples to be measured by Raman
spectroscopy. Samples for TEM are prepared following a similar
process to that previously utilized to make free-standing and
TEM-compatible carbon nanotube devices~\cite{Meyer2006}.
\begin{figure}[!tb]
\includegraphics[width=\columnwidth]{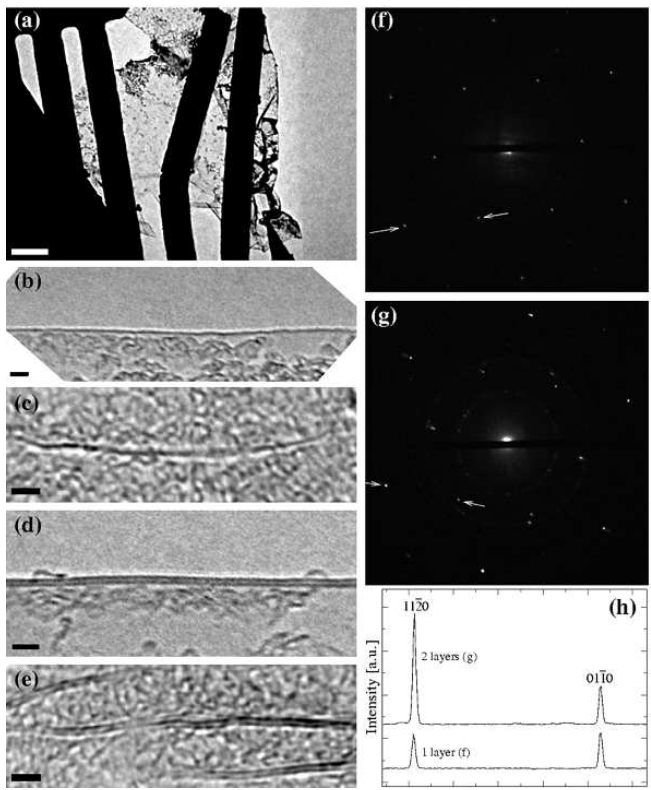}\\
\caption{(a) TEM image of a suspended graphene sheet. The metal
grid is also visible in the optical microscope. (b) High
resolution image of a folded edge of single layer graphene and (c)
a wrinkle within the single layer sheet. (d) Folded edge of a
two-layer sample and (e) internal foldings of the two-layer sheet.
The amorphous contrast on the sheets is most likely due to
hydrocarbon adsorbates on the samples that were cracked by the
electron beam. (f) Electron diffraction pattern for close to
normal incidence from single layer graphene and (g) from two
layers. (h) Intensity profile plot along the line indicated by the
arrows in (f+g). The relative intensities of the spots in the
two-layer sheet are consistent only with A-B (and not A-A)
stacking. Scale bars: (a) 500 nm; (b-e) 2 nm.}\label{Fig1}
\end{figure}
In addition, this allows us to have free-standing layers on a grid
easily seen in the optical Raman microscope, facilitating their
location during Raman measurements, Fig.~\ref{Fig1}(a). Electron
diffraction is done in a Zeiss 912$\Omega$ microscope at a voltage
of 60kV, and high-resolution images are obtained with a Philips
CM200 microscope at 120kV. A HR-TEM analysis of foldings at the
edges or within the free hanging sheets gives the number of layers
by direct visualization, since at a folding the sheet is locally
parallel to the beam, Fig.~\ref{Fig1}(b-e). Edges and foldings of
the one or two layers are dominated by one or two dark lines. The
number of layers is also obtained by a diffraction analysis of the
freely suspended sheets for varying incidence angles, and confirms
the number of layers seen in the foldings, Fig.~\ref{Fig1}(d,e). In
particular, the diffraction analysis of the bi-layer shows that it
is A-B stacked (the intensity of the 11-20 diffraction spots (outer
hexagon) is roughly twice that of the 1-100 (inner hexagon),
Fig.~\ref{Fig1}(h), in agreement with image simulations. This
confirms that multi-layer graphene maintains the same stacking as
graphite.

Unpolarized Raman spectra are measured on single, bi and
multi-layers on Si+SiO$_2$. Some are then processed into
free-hanging sheets, also measured by TEM as described above, and
measured again by Raman spectroscopy after TEM. The measurements are
performed at room temperature with a Renishaw spectrometer at 514
and 633 nm. A 100$\times$ objective is used. Extreme care is taken
to avoid sample damage or laser induced heating. Measurements are
performed from $\sim4$mW to $\sim0.04$mW incident power. No
significant change in the spectra is observed in this power range
both for free standing and supported samples. The Raman spectra of
suspended and on-substrate graphene are similar, one of the main
differences being a D peak observed for the much smaller samples
used for TEM. We also measure the reference bulk graphite used to
produce the layers.
\begin{figure}[htp]
\centering
\includegraphics[width=80mm]{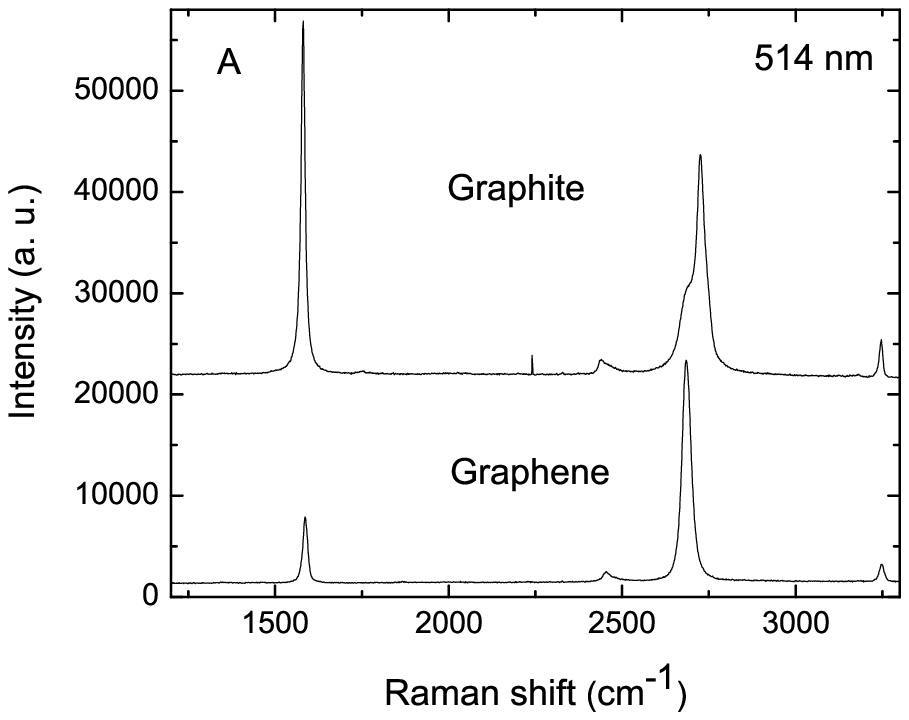}\\
\includegraphics[width=70mm]{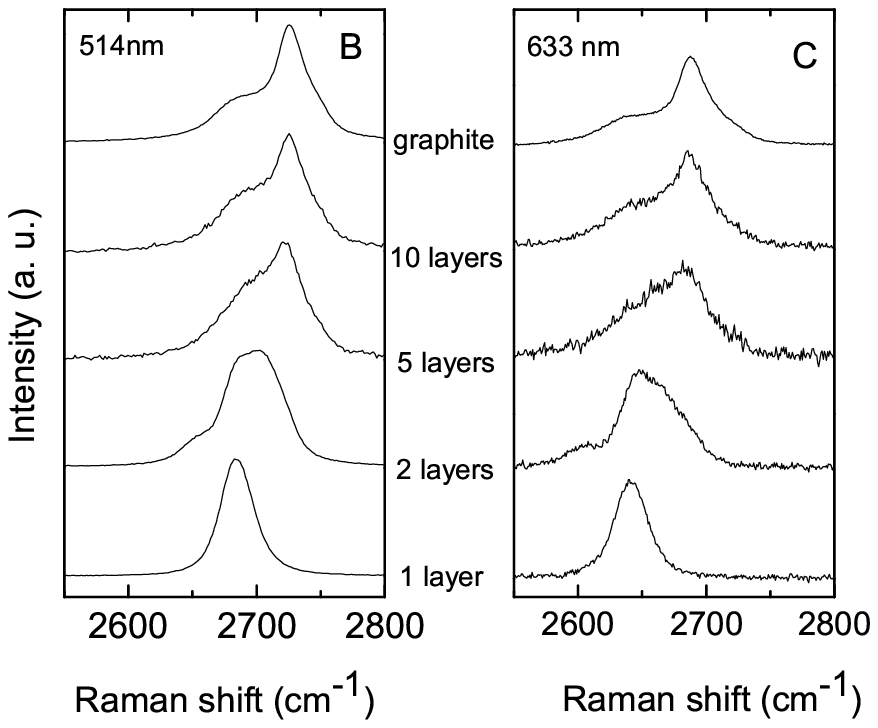}\\
\includegraphics[width=40mm]{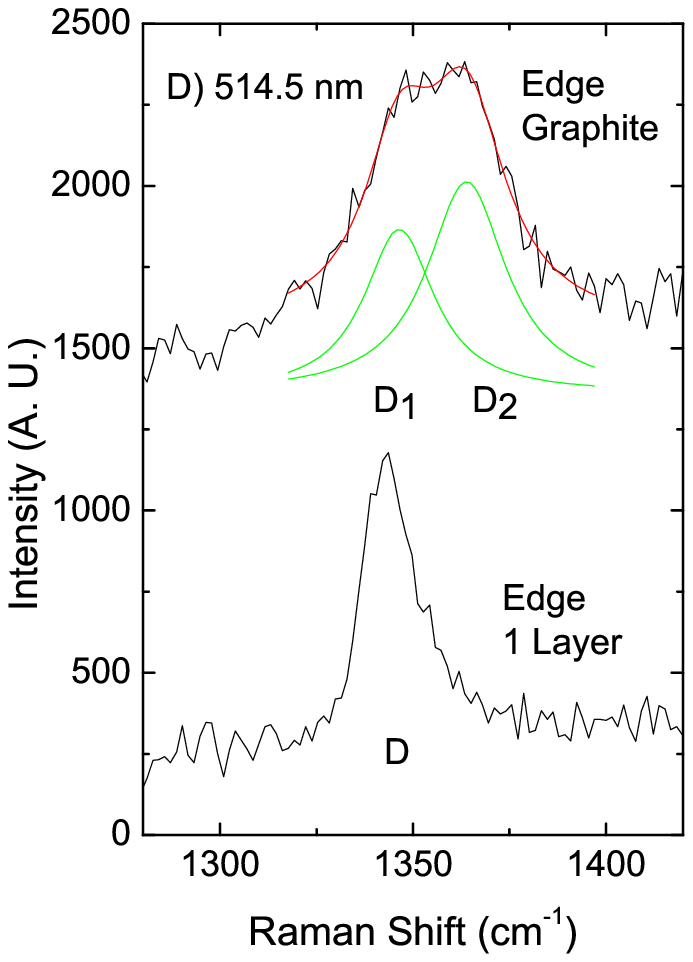}
\includegraphics[width=40mm]{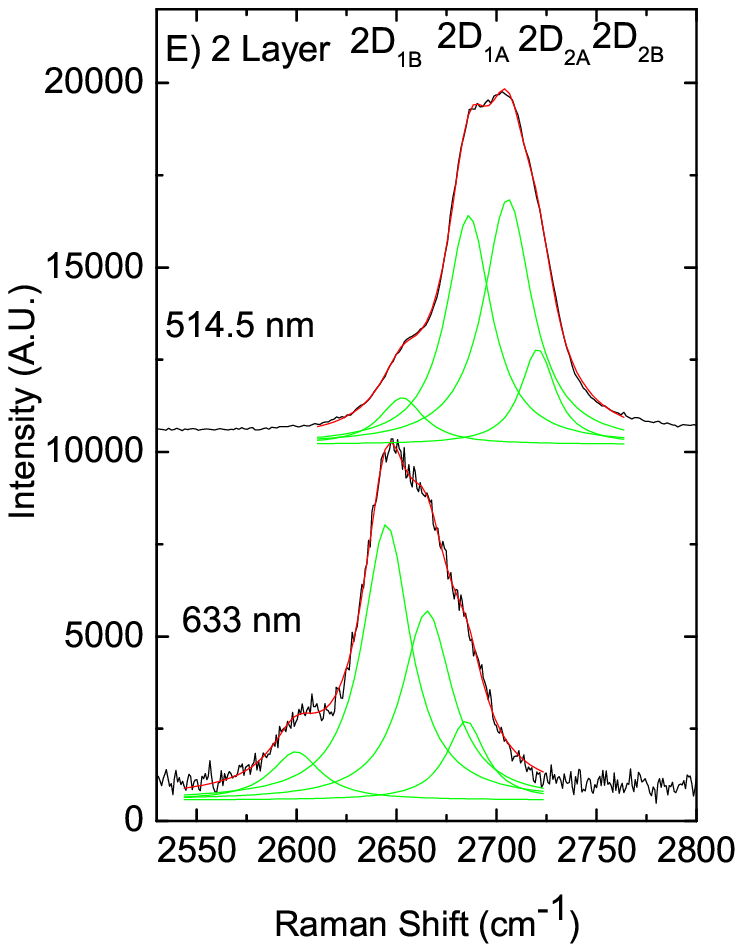}\\
\caption{(a) Comparison of Raman spectra at 514 nm for bulk
graphite and graphene. They are scaled to have similar height of
the 2D peaks. (b) Evolution of the spectra at 514 nm with the
number of layers. (c) Evolution of the Raman spectra at 633 nm
with the number of layers. (d) Comparison of the D band at 514 nm
at the edge of bulk graphite and single layer graphene. The Fit of
the D$_1$ and D$_2$ components of the D band of bulk graphite is
shown. (e) The four components of the 2D band in 2 layer graphene
at 514 nm and 633 nm.}\label{Fig2}
\end{figure}

Fig.~\ref{Fig2}(a) compares the 514 nm Raman spectra of graphene and
bulk graphite. The two most intense features are the G peak at
$\sim1580$cm$^{-1}$ and a band at $\sim2700$cm$^{-1}$, historically
named G', since it is the second most prominent band always observed
in graphite samples\cite{Vidano1981}. The G peak is due to the
doubly degenerate zone centre $E_{2g}$ mode~\cite{ Tuinstra1970}. On
the contrary, the G' band has nothing to do with the G peak, but is
the second order of zone boundary phonons. Since zone-boundary
phonons do not satisfy the Raman fundamental selection rule, they
are not seen in the first order Raman spectra of defect-free
graphite~\cite{ Nemanich1979}. Such phonons give rise to a Raman
peak at $\sim1350$cm$^{-1}$ in defected graphite, called D
peak~\cite{ Tuinstra1970}. Thus, for clarity, we refer to the G'
peak as 2D. Fig.~\ref{Fig2}(a) shows that no D peak is observed in
the centre of the graphene layers. This proves the absence of a
significant number of defects in the structure. As expected, a D
peak is only observed at the sample edge, Fig.~\ref{Fig2}(d).
Fig.~\ref{Fig2}(a) shows a significant change in the shape and
intensity of the 2D peak of graphene compared to bulk graphite. The
2D peak in bulk graphite consists of two components 2D$_{\rm 1}$ and
2D$_2$~\cite{ Vidano1981, Nemanich1979}, roughly $1/4$ and $1/2$ the
height of the G peak, respectively. Here we measure a single, sharp
2D peak in graphene, roughly 4 times more intense than the G peak.
Notably, the G peak intensity of single layer and bulk graphite is
comparable (note that Fig.~\ref{Fig2}(a) is re-scaled to show a
similar 2D intensity) and the G position is 3-5 cm$^{-1}$ higher
than bulk graphite. The change in shape of the 2D band is nicely
confirmed in Fig.~\ref{Fig2}(d), which compares the D peak observed
on the graphite edge with that of the graphene edge. The graphene D
peak is a single sharp peak, while that of graphite is a band
consisting of two peaks D$_{\rm 1}$and D$_{\rm 2}$\cite{Vidano1981}.
Fig.~\ref{Fig2}(b,c) plot the evolution of the 2D band as a function
of the number of layers for 514.5 nm and 633 nm excitations. These
immediately indicate that bi-layer graphene has a much broader and
up-shifted 2D band with respect to graphene. This band is also quite
different from bulk graphite. It has 4 components, 2D$_{\rm 1B}$,
2D$_{\rm 1A}$, 2D$_{\rm 2A}$, 2D$_{\rm 2B}$, 2 of which, 2D$_{\rm
1A}$ and 2D$_{\rm 2A}$, have higher relative intensities than the
other 2, as indicated in Fig.~\ref{Fig2}(e). Fig.~\ref{Fig2}(b,c)
show that a further increase of the number of layers leads to a
significant decrease of the relative intensity of the lower
frequency 2D$_{\rm 1}$ peaks. For more than 5 layers the Raman
spectrum becomes hardly distinguishable from that of bulk graphite.
Thus Raman spectroscopy can clearly identify a single layer, from
bi-layer from few (less than 5) layers. This also explains why
previous experiments on nano-graphites, but not individual or
bi-layer graphene, failed to identify these features
\cite{Cancado2002, Cancado2004}. In particular, it was noted from
early studies that turbostratic graphite (i.e. without AB stacking)
has a single 2D peak~\cite{Lespade1984}. However, its Full Width at
Half Maximum (FWHM) is ~50 cm$^{-1}$ almost double that of the 2D
peak of graphene and upshifted of ~20 cm$^{-1}$. Turbostratic
graphite also often has a first order D peak~\cite{Lespade1984}.
SWNTs show a sharp 2D peak similar to that we measure here for
graphene~\cite{Jorio2002bis}. The close similarity (in position and
FWHM) of our measured graphene 2D peak and the 2D peak in SWNTs of
1-2 nm diameter~\cite{ Souza2003} implies that curvature effects are
small for the 2D peak for SWNTs in this diameter range, the most
commonly found in experiments. This questions the assumption that
the 2D peak in SWNT should scale to the up-shifted average 2D peak
position in bulk graphite for large diameters \cite{ Souza2003}.
This assumption was utilized to fit a scaling law relating SWNT
diameter and 2D peak position, which is often used to derive the
diameter of inner tubes in double wall nanotubes~\cite{Souza2003,
Pfeiffer2005}. Despite the similarities, it is important to note
that there are major differences between graphene and SWNT Raman
spectra, which allow to easily distinguish these materials. Indeed,
confinement and curvature split the two degenerate modes of the G
peak in SWNTs~\cite{ Jorio2002bis}, resulting in G$^+$ and G$^-$
peaks.

We now explain why graphene has a single 2D peak, and why this
splits in four components in bi-layer graphene. Several authors
previously attempted to explain the double structure of the 2D
peak in graphite~\cite{Cancado2004, Vidano1981, Nemanich1979,
Lespade1984, Cancado2002, Maultzsch2004prb}, however they always
neglected the evolution of the electronic bands with the number of
layers, which is, on the contrary, the key fact. The 2D peak in
graphene is due to two phonons with opposite momentum in the
highest optical branch near the \textbf{K} ($A'_1$ symmetry at
\textbf{K})~\cite{ Ferrari2000, Piscanec2004, Tuinstra1970}.
Fig.~\ref{Fig2} shows that this peak changes in position with
varying excitation energy. This is due to a Double Resonance (DR)
process, which links the phonon wave-vectors to the electronic
band structure~\cite{ Thomsen2000}.
\begin{figure}[!tb]
\includegraphics[width=65mm]{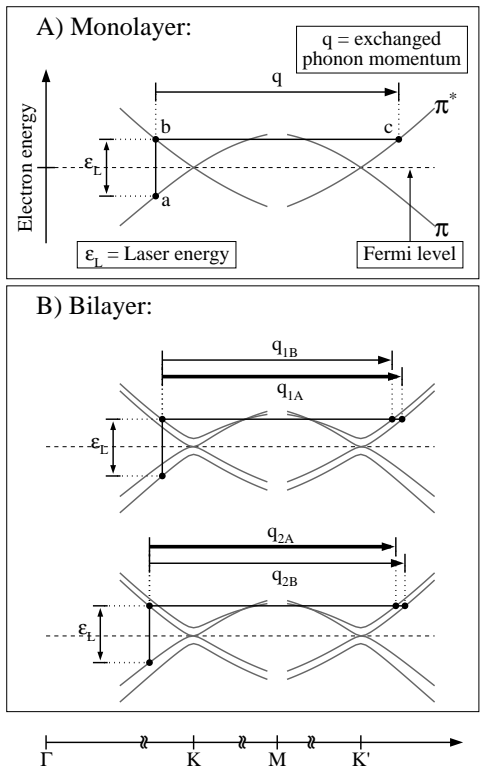}\\
\caption{DR scheme for the 2D peak in (a) single layer and (b)
bi-layer graphene. }\label{Fig3}
\end{figure}
Within DR, Raman scattering is a third order process involving
four virtual transitions: i) a laser induced excitation of an
electron/hole pair (a$\to$b vertical transition in
Fig.~\ref{Fig3}(a)); ii) electron-phonon scattering with an
exchanged momentum $\bf q$ close to $\bf K$ (b$\to$c); iii)
electron-phonon scattering with an exchanged momentum $\bf -q$
(c$\to$b); iv) electron/hole recombination (b$\to$a). The DR
condition is reached when the energy is conserved in these
transitions.  The resulting 2D Raman frequency is twice the
frequency of the scattering phonon, with $\bf q$ determined by the
DR condition. For simplicity, Fig.~\ref{Fig3}(a,b) neglect the
phonon energy and do not show the equivalent processes for
hole-phonon scattering.

Consistent with the experimental observation of a single component
for the 2D peak in single layer graphene, Fig.~\ref{Fig3}(a,b)
only shows the phonon satisfying DR conditions with momentum
q$>$K, along the $\bm \Gamma-\textbf{K}-\textbf{M}$ direction
(K$<$q$<$M ).
\begin{table}
\begin{ruledtabular}
\begin{tabular}{c|cccc|cc}
{\bf 514.5} & \multicolumn{4}{c|}{2 Layers}\\
\hline {\bf Experimental} & -44 & -10 & +10 & +25  \\
       {\bf Theory}       & -44 & -11 & +11 & +41  \\
\hline {\bf 633}          &     &     &     &      \\
\hline {\bf Experimental} & -55 & -10 & +10 & +30  \\
       {\bf Theory}       & -44 &  -9 &  +9 & +41  \\
\end{tabular}
\end{ruledtabular}
\caption{Relative splitting of 2D components in bi-layer graphene.
In each case, we show the shift with respect to the average
frequency of the two main peaks. The four columns of the bi-layer
correspond to processes q$_{1B}$, q$_{1A}$, q$_{2A}$, q$_{2B}$,
respectively. The theoretical values are obtained by multiplying the
DR q vectors determined from the DFT electronic bands by dw/dq= 645
cm$^{-1}$ \AA. Here dw/dq is the ratio between the measured shift of
the 2D peak frequency with the laser energy in graphene ($\sim$ 99
cm$^{-1}$/eV), and the corresponding variation of the DR q vector
computed from the DFT electronic bands.}\label{tab1}
\end{table}
The other two possible DR phonons, with q$<$K and q$\sim$K, give a
much smaller contribution to the Raman intensity. In fact, the q$<$K
phonon involves a smaller portion of the phase-space because of the
band-structure trigonal warping (see Fig.4 of Ref.\cite{ Kurti2002}
and related discussion) and the q$\sim$K phonon has a zero
electron-phonon coupling for this transition, as discussed in Ref
\cite{ Piscanec2004} (see footnote 24, for q$\sim$K, $\theta''=0$)
and Ref. \cite{ Maultzsch2004prb}. This differs from the models of
Ref.~\cite{Cancado2002, Maultzsch2004prb}, which predict 2 similar
components for the D peak even in single layer, in disagreement with
the experiments of Fig.~\ref{Fig2}.

We now examine the bi-layer case. The observed 4 components of the
2D peak could in principle be attributed to two different
mechanisms: the splitting of the phonon branches~\cite{ acfbook,
Vidano1981, Nemanich1979, Lespade1984}, or the spitting of the
electronic bands~\cite{Ferrari2000}. To ascertain this we compute
the phonon frequencies \cite{Piscanec2004} for both single and
bi-layer graphene (stacked AB, as indicated by TEM), at the q
corresponding to the DR condition for the 514 and 633 nm lasers.
The splitting of the phonon branches is $<$1.5 cm$^{-1}$, much
smaller than the experimentally observed 2D splitting. Thus, this
is solely due to electronic bands effects. In the bi-layer, the
interaction of the graphene planes causes the $\pi$ and $\pi^*$
bands to divide in four bands, with a different splitting for
electrons and holes, Fig.~\ref{Fig3}(b). Amongst the 4 possible
optical transitions, the incident light couples more strongly the
two transitions shown in Fig.~\ref{Fig3}(b). The two almost
degenerate phonons in the highest optical branch couple all
electron bands amongst them. The resulting four processes involve
phonons with momenta q$_{\rm 1B}$, q$_{\rm 1A}$, q$_{\rm 2A}$, and
q$_{\rm 2B}$, as shown in Fig.~\ref{Fig3}(b). The four
corresponding processes for the holes, and those associated to the
2 less intense optical transitions [not shown in
Fig.~\ref{Fig3}(b)], are associated to momenta almost identical to
q$_{\rm 1B}$, q$_{\rm 1A}$, q$_{\rm 2A}$, q$_{\rm 2B}$. These
wave-vectors correspond to phonons with different frequencies, due
to the strong phonon dispersion around \textbf{K} induced by the
electron-phonon coupling~\cite{ Piscanec2004}. They produce four
different peaks in the Raman spectrum of bi-layer graphene.
Tab.~\ref{tab1} reports the expected splittings and shows that
they compare very well with experiments.

In conclusion, graphene's electronic structure is uniquely
captured in its Raman spectrum, that clearly evolves with the
number of layers. Raman fingerprints for single-, bi- and
few-layer graphene reflect changes in the electronic structure and
electron-phonon interactions and allow unambiguous,
high-throughput, non-destructive identification of graphene
layers.

A.C.F. acknowledges funding from EPSRC GR/S97613, The Royal
Society and The Leverhulme Trust; C.C. from the Oppenheimer Fund.

\end{document}